\documentclass[aps,pre,twocolumn,showpacs,floatfix,preprintnumbers,amsmath,amssymb]{revtex4}
\usepackage{epsfig}
\usepackage{graphicx}
\usepackage{dcolumn}
\usepackage{bm}

\begin{document}
\title{Return interval distribution of extreme events and long term memory}

\author{M. S. Santhanam\footnote{Permanent address:
Physical Research Laboratory, Navrangpura, Ahmedabad 380 009, India.} and Holger Kantz}

\affiliation{Max Planck Institute for the Physics of Complex Systems,\\
N\"othnitzer Strasse 38., Dresden 01187, Germany.}

\date{\today}

\begin{abstract}
The distribution of recurrence times or return intervals between extreme events
is important to characterize and understand the behavior of physical
systems and phenomena in many disciplines. It is well known that many physical
processes in nature and society display long range correlations. Hence,
in the last few years, considerable
research effort has been directed towards studying the distribution of
return intervals for long range correlated time series. Based on numerical
simulations, it was shown that the return interval distributions are
of stretched exponential type. In this paper, we obtain an analytical
expression for the distribution of return intervals in long range correlated
time series which holds good when the average return intervals are large.
We show that the distribution is actually a product of power law and a
stretched exponential form. We also discuss the regimes of validity and perform
detailed studies on how the return interval distribution depends on
the threshold used to define extreme events.
\end{abstract}

\maketitle

\section{Introduction}
  Extreme events take place frequently in both nature and society. For instance,
the recurrence of floods, droughts, earth quakes and economic recession
are all examples of extreme events. The consequences of extreme events to life and
property are often enormous and hence it is desirable to study their
properties and questions related to their predictability.
Interestingly, all of these extreme events are also non-equilibrium phenomena and
studying the extreme value statistics in them will lead to a better
understanding of the models and the phenomenology of non-equilibrium statistical physics.
Thus, there is an increasing interest in the physics literature to understand a
broad range of issues and phenomena connected with the occurance of extreme events
and their dynamics \cite{kantz1,reiss}. 

In the classical extreme value theory,
the limiting distribution for the extreme maximal values
in sequences of independent and identically distributed random variables
can be one of the Fr\'{e}chet, Gumbel or Weibull distribution depending
on the behavior of the tail of the probability density \cite{gumbel}.
This has been empirically verified in many cases of practical interest.
Many new applications continue to be discovered, for example, the
recent one being the distribution of extreme components of the eigenmodes of
quantum chaotic systems \cite{arul}.
In contrast to the questions about the distribution of extrema, one of the
problems being addressed in the last few years is
the distribution of the returns intervals for the extreme events when the
underlying time series displays long memory \cite{long1,long2,long3,long4,long5}.
This is primarily motivated by
the fact that many of the natural and socio-economic phenomena, e.g., 
daily temperature, DNA sequences, river run-off, earth quakes, stock markets etc.,
display long memory or long range correlation \cite{lmem,lmem1}. Long memory
implies slowly decaying auto correlation function of the power law
type such that the system does not exhibit typical time scales.
In this case, the intervals between extreme events are likely to be correlated
as well.
On the contrary, it is known that for an uncorrelated time series, intervals between
extreme events are also uncorrelated and are exponentially distributed.
The question is how the presence of long range correlation modifies
the return interval distribution of extreme events ? A definite answer to
this question would shed new light on many problems across various disciplines.

Return interval distributions are interesting and useful for several reasons, the
most important being that many problems in diverse fields can be formulated
in terms of return interval statistics with wide ranging applications.
For instance, the problem of recurrence
time interval between earthquakes above a given magnitude \cite{corral},
x-ray solar flare recurrences \cite{wland}, statistics of acoustic emission
from rock fractures \cite{david}, inter arrival packet times on computer
and cellular networks \cite{ant} and the classical problem of Poincare
recurrences in Hamiltonian systems \cite{edu} can all be formulated
as extreme event questions involving return interval distribution.
In a non-stationary time series, it is
often difficult to reliably estimate its temporal statistical properties
such as the autocorrelations or higher order correlations.
Thus, return interval distributions are also a useful tool to characterize
temporal properties of such systems.

\begin{figure}
\includegraphics*[width=6cm]{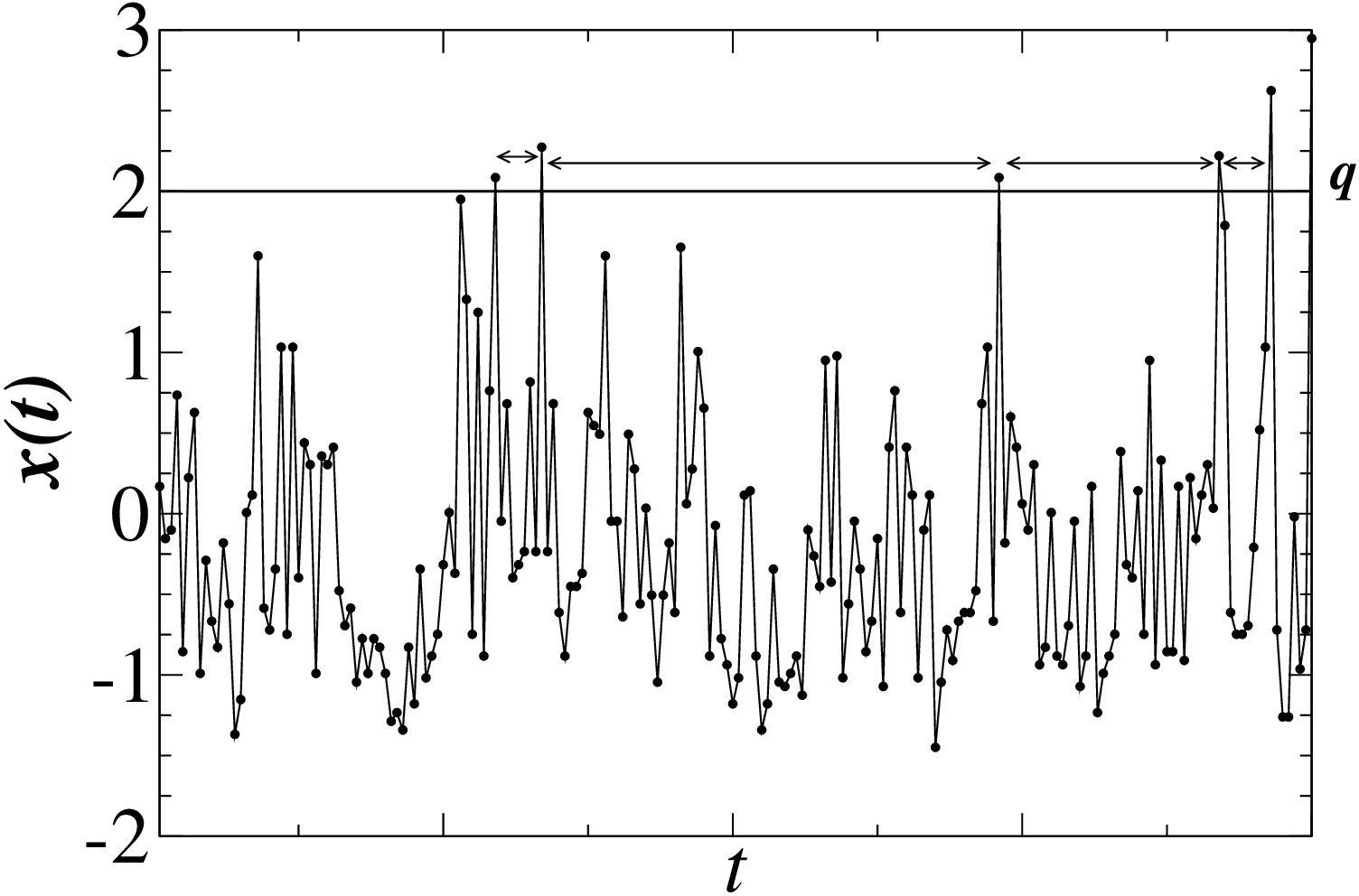}
\caption{This schematic diagram shows the return intervals for
a threshold value $q=2$ as a function of time $t$.}
\label{scheme}
\end{figure}

  Let $x(t)$ denote a sequence of random variable, where $t$ is the time index.
We will call an event extreme if $x(t) > q$ where $q$ is some threshold value.
The return interval $r$ is the time between successive occurance of extreme events.
With respect to threshold $q$, we have a well-defined series of return intervals,
$r_k$, $k=1,2,3,...N$. This is schematically shown in Fig \ref{scheme}.
If the random variables $x(t)$ are uncorrelated, then the return intervals $r_k$ are 
also uncorrelated and they are exponentially distributed as
\begin{equation}
P_q(r) = \frac{1}{\langle r \rangle} \; e^{-r/\langle r \rangle}.
\end{equation}
In order to use later, we also define the average return interval dependent on
threshold $q$ to be
\begin{equation}
\langle r \rangle_q  = \lim_{N\to \infty} \frac{1}{N} \sum_{k=1}^N r_k.
\end{equation}
In contrast to an uncorrelated time series, a long range correlated series
has an autocorrelation function that displays power law of the form,
\begin{equation}
C(\tau) = \langle x(t+\tau) ~x(t) \rangle \sim \tau^{-\gamma}, \;\;\;\;\;\;\;\;\;\;\;
(0 < \gamma < 1),
\end{equation}
where $\langle . \rangle$ denotes the temporal average and $\gamma$ is the
auto correlation exponent.
The work done in the last few years show that the long range correlation
does indeed affect the return interval distribution of extreme
events \cite{long1,long2,long3,long4,long5}.
Empirical results in a series of papers \cite{long1,long2,long3,long4,long5}
have shown that, in the presence
of long range correlation, the return interval distribution becomes a
stretched exponential given by,
\begin{equation}
P_q(R) = A(\gamma) ~ e^{-B(\gamma) ~R^{\gamma} }.
\label{rtd1}
\end{equation}
with scaled return intervals being defined as $R=r/\langle r \rangle$.
Both $A(\gamma)$ and $B(\gamma)$ are constants that depend on $\gamma$.
They can be fixed by normalizing both the probability
and the average return interval to unity. It has also been shown that
the return intervals themselves are long range correlated.

However, an analytical justification for the stretched exponential
distribution in Eq. \ref{rtd1} is still lacking and the main
contribution of this paper is to partly fill this void.
In this context, it must be noted that deviations from the stretched
exponential distribution in Eq. \ref{rtd1} have been noted for
return intervals shorter ($R<1$) than the average. For short
return intervals, i.e, $R<1$, empirical results display a
power law with the exponent $\sim (\gamma-1)$ \cite{long3}, which is not
explained by Eq. \ref{rtd1}. While the return interval distribution is expected
to depend on the threshold $q$, the stretched exponential form does
not explicitly reveal this dependence. This paper addresses
these questions using a combination of analytical and numerical
results. Firstly, from theoretical arguments, we obtain an
approximate expression for the return interval distribution, which
modifies Eq. \ref{rtd1} from a purely stretched exponential form to a
product of power law and stretched exponential. Secondly, we systematically study the
dependence of return interval distribution on the threshold $q$ and show
that our analytical result holds good in the limit of $q >> 1$. In general,
the return interval distribution depends on the value of threshold $q$.

Recently, in the study of global seismic activity above some
magnitude $M$, the distribution $F(\tau)$ of recurrence
times $\tau$ have been shown to follow a scaling ansatz of the form
\begin{equation}
F(\tau) = \frac{1}{\bar{\tau}} f(\tau/\bar{\tau}),
\end{equation}
where the function $f(\tau/\bar{\tau})$ is the
gamma distribution \cite{corral}. In fact, this scaling relation
seems to hold good for forest fire occurance intervals \cite{corral1},
tsunami inter event times \cite{geist} and ion channel currents in
voltage dependent anion channels in the cell \cite{mkv}. The analytical
distribution obtained in this paper might shed light on this
scaling found in a variety of systems.

In the next section, we obtain an analytical expression
for the return interval distribution and in the subsequent section
we present our numerical results. Further, we systematically study
the dependence of the return interval distribution on the threshold
used to define the extreme event. Finally, we present our conclusions.

\section{Return time distribution}
The starting point of our approach is to transform the given long range correlated
time series $x(t)$ with autocorrelation exponent $\gamma$ into a 
binary sequence with 1 at positions of
extreme events and 0 elsewhere. Thus, we obtain new binary
sequence defined by,
\begin{eqnarray}
y(t) & = & 1, ~~~\mbox{if}~~~ x(t) \ge q \nonumber \\
     & = & 0, ~~~\mbox{if}~~~ x(t) < q
\label{threshold}
\end{eqnarray}
We use the empirical result that for a long range
correlated time series with the autocorrelation exponent $\gamma$,
the return intervals are also long range correlated with the
same exponent. 
Thus, our probability model is the statement that
given an extreme event at time $t=0$, the probability to find an
extreme event at time $t=r$ is given by,
\begin{equation}
P_{ex}(r) = a r^{-(2H-1)} = a r^{-(1-\gamma)},
\label{probmodel}
\end{equation}
where $1/2 < H < 1$ is the Hurst exponent \cite{hurst} and
$a$ is the normalization constant that
will be fixed later. We have also used the well-known relation between Hurst exponent
and autocorrelation exponent;  $\gamma = 2-2H$. Equation \ref{probmodel} implies
that after an extreme event it is highly probable to expect the next event
to be an extreme one too; and this is a reasonable proposition for a
persistent time series. Notice also that for an uncorrelated time series
$H=1/2$. This leads to $P(r)$ in Eq. \ref{probmodel} becoming independent
of $r$, as would be expected for an uncorrelated time series.
Further support for this proposition comes from the theorem due to
Newell and Rosenblatt \cite{newell,currsci} obtained in the context of zero
crossing probabilities
for Gaussian processes. It states that for a separable Gaussian
stationary process $X(t)$ with mean $\langle X \rangle =0$, the probability
$g(T)$ that $X(t) >0$ for $0 \ge t \ge T$
is $g(T) = O(T^{-\alpha})$ as $T \to \infty$, where $\alpha > 0$.

Next we calculate the probability that given an extreme event at
time $t=0$, no extreme event occurs in the interval $(0,r)$.  For this,
we divide the interval $r$ into $m$ sub-intervals indexed by $j=0,1,2.....(m-1)$
and we calculate
this probability in each of the intervals. For the $j$th sub-interval,
using Eq. \ref{probmodel}, the probability of extreme event is given by,
\begin{eqnarray}
h(j) & = & \frac{a~r}{m} \left( \frac{(j+1)r}{m} \right)^{-(1-\gamma)} +  \\
     &   &   \frac{a~r}{2m} \left[ \left(\frac{jr}{m} \right)^{-(1-\gamma)}
        - \left(\frac{(j+1)r}{m} \right)^{-(1-\gamma)} \right]  \nonumber
\end{eqnarray}
After simplifying this expression, the probability that no extreme event
occurs in the $j$th sub-interval is given by,
\begin{equation}
1-h(j) = 1 - \frac{ar}{2m} \left( \frac{r}{m} \right)^{-(1-\gamma)}
  \left[ (j+1)^{-(1-\gamma)} + j^{-(1-\gamma)} \right]
\end{equation}
At this point, we make an approximation and assume that the
probability of no extreme event occurance in each sub-interval to be 
an independent event. Then, the probability $P_{noex}(r)$ that no extreme event occurs
in any of the $m$ sub-intervals in $(0,r)$ is simply the
product of probabilities,
\begin{equation}
P_{noex}(r) = \lim_{m\to\infty} \prod_{j=0}^{m-1} ~ 1-h(j).
\end{equation}

The required probability $P(r)~dr$ is simply the product of $P_{noex}$  with
the probability $P_{ex}$ that an extreme event takes place in the infinitesimal
interval $dr$ beyond $r$. This can be assembled together as,
\begin{eqnarray}
P(r)~dr & = & P_{noex}(r) ~P_{ex}(r) ~dr \nonumber \\
        & = & \lim_{m\to\infty} \left[1-\phi_{m,r} \right]
                                \left[1-\phi_{m,r} (2^{-\gamma}+1) \right] \label{prob1} \\
        &   &                  \left[1-\phi_{m,r} (3^{-\gamma}+2^{-\gamma}) \right] \ldots\ldots \nonumber \\
        &   &  \left[1-\phi_{m,r}  (m^{-\gamma}+(m-1)^{-\gamma}) \right] ~ a ~r^{-(1-\gamma)} dr \nonumber
\end{eqnarray}
where,
\begin{equation}
\phi_{m,r} = \frac{a}{2}\left(\frac{r}{m}\right)^{-\gamma}
\label{short1}
\end{equation}
The value of $m$ can be arbitrarily large and the Eq. \ref{prob1} can be
simplified and rewritten as,
\begin{eqnarray}
P(r)~dr & = & \lim_{m\to\infty} \exp\left(-\frac{a}{2} \left( \frac{r}{m} \right)^\gamma 
              \{ 2 H_{m-1}^{(\gamma-1)} + \right. \nonumber \\
        &   &  \left. m^{-(1-\gamma)}   \}    \right) a ~r^{-(1-\gamma)} dr
\label{prob2}
\end{eqnarray}
where $H_{m-1}^{(\gamma-1)}$ is the generalized Harmonic number \cite{harmonic}.
In order to take the limit $m\to\infty$, we note that
\begin{equation}
\lim_{m\to\infty} \frac{H_{m-1}^{(\gamma-1)}}{m^\gamma} = \frac{1}{\gamma},
\;\;\;\;\;\;\;\;\;\; (0 < \gamma < 1).
\label{limh}
\end{equation}
Using this Eq. \ref{limh} in Eq. \ref{prob2} and taking the limit, we obtain
the following result for the distribution of return intervals;
\begin{equation}
P(r) dr =  a ~r^{-(1-\gamma)} ~e^{-\frac{a}{\gamma} r^{\gamma}} dr .
\label{rtd2}
\end{equation}
The constant $a$ will be fixed by normalization as follows ;
we demand that the total probability and the average return interval 
$\langle r \rangle$ be normalized to unity.
\begin{eqnarray}
I  & = & \int_0^{\infty} ~P(r) ~dr = 1   \;\;\;\;\;\;\;\;\; \mbox{and} \label{normint0} \\
\langle r \rangle & = & \int_0^{\infty} ~r ~P(r) ~dr = 1
\label{normint}
\end{eqnarray}
However, the distribution in Eq. \ref{rtd2} is already normalized and hence
Eq. \ref{normint} will be used to determine the value of $a$.
The requirement that
$\langle r \rangle = 1$ is equivalent to transforming the return
intervals $r$ in units of $\langle r \rangle$.
Performing the integrals above, the normalized distribution in the
variable $R=r/\langle r \rangle$ turns out to be,
\begin{equation}
P(R) = \gamma \left[ \Gamma\left(\frac{1+\gamma}{\gamma}\right) \right]^{\gamma} ~R^{-(1-\gamma)}
~e^{-\left[ \Gamma\left(\frac{1+\gamma}{\gamma}\right) \right]^{\gamma} R^{\gamma}}
\label{rtd3}
\end{equation}
where $\Gamma(.)$ is the Gamma function.
First, we discuss some of the salient features of this distribution.
The case $\gamma=1$ defines the crossover to short range or uncorrelated
time series.
If we put $\gamma=1$ in the distribution in Eq. \ref{rtd3} above, we recover
the exponential distribution, $P(R) = \exp(-R)$.
In the region $R \ll 1$, i.e., for the return intervals much below the
average, the dominant behavior can be seen by taking logarithm on
both sides of Eq. \ref{rtd3} leading to,
\begin{equation}
\log P(R) = \log(\gamma ~g_{\gamma}) -(1-\gamma)  \log R - g_{\gamma} ~ R^\gamma,
\label{logp}
\end{equation}
where we have used
$g_{\gamma} = \left[ \Gamma\left(\frac{1+\gamma}{\gamma}\right) \right]^{\gamma}$.
For $R \ll 1$, the second term dominates the distribution and thus we
obtain a power law with an exponent $(\gamma-1)$; 
\begin{equation}
P(R) \propto R^{-(1-\gamma)}   \;\;\;\;\;\;\;\;\;\;  (R \ll 1)
\label{power}
\end{equation}
This power law behavior with exponent $(\gamma-1)$ for short return intervals
has already been noted in the
numerical results presented in Ref. \cite{long3}. Thus, our approach
analytically shows the emergence of a power law regime for short $R$ in contrast to
the stretched exponential distribution.
On the other hand, for $R \gg 1$, the logarithmic
term in Eq. \ref{logp} can be dropped and the return interval distribution 
behaves essentially like a stretched exponential distribution,
\begin{equation}
P(R) \propto e^{- g_{\gamma} ~ R^\gamma}.   \;\;\;\;\;\;\;\;\;\;  (R \gg 1)
\end{equation}
Thus, stretched exponential is a good approximation for $R \gg 1$.
This partly explains why a pure stretched exponential distribution as in
Eq. \ref{rtd1} deviates, for $R<1$, from the simulated return interval
distributions in the earlier works \cite{long1,long2,long3,long4,long5}.
Finally, we also note that Eq. \ref{rtd3} can also be derived by other
methods without actually discretising the interval $r$ as we have done.

As shown above, the return interval distribution in Eq. \ref{rtd3} does
reproduce the empirical results already known in the literature but
is nevertheless approximate in
the following sense. It is known that there exist correlations among
the return intervals and they are particularly strong as $\gamma \to 0$.
Thus, every return interval depends on the value of previous
return interval. This is also well documented in the literature
as the conditional probability $P(R|R_0)$ to find return interval $R$,
given that the previous return interval was $R_0$ \cite{long3,long4,long5}.
This conditional
probability shows interesting features and deviates from the
case of uncorrelated return intervals. Equation \ref{rtd3} does not
take into account these correlations among intervals and in fact is
derived on the assumption that return intervals are independent.
This is a gross approximation though in the absence any other
definitive model for the correlations among intervals this is
a simple and analytically tractable choice. Based on this argument, one can
expect Eq. \ref{rtd3} to describe the return interval statistics
in the regime where the correlations are not highly dominant,
for $\langle r \rangle \gg 1$ \cite{berman}. Secondly, note that
even though threshold $q$ plays a crucial role as we will describe in
the next section, it does not play any role in Eq. \ref{rtd3}.
Threshold $q$ is related to $\langle r \rangle$ such that higher the value
of $q$, larger is $\langle r \rangle$, though it is not a linear relation. 
Thus, the theoretical arguments
leading to Eq. \ref{rtd3} would best describe an asymptotic limit
of $q \gg 1$ or $\langle r \rangle \gg 1$.

\begin{widetext}
\begin{figure}
\includegraphics*[width=14cm]{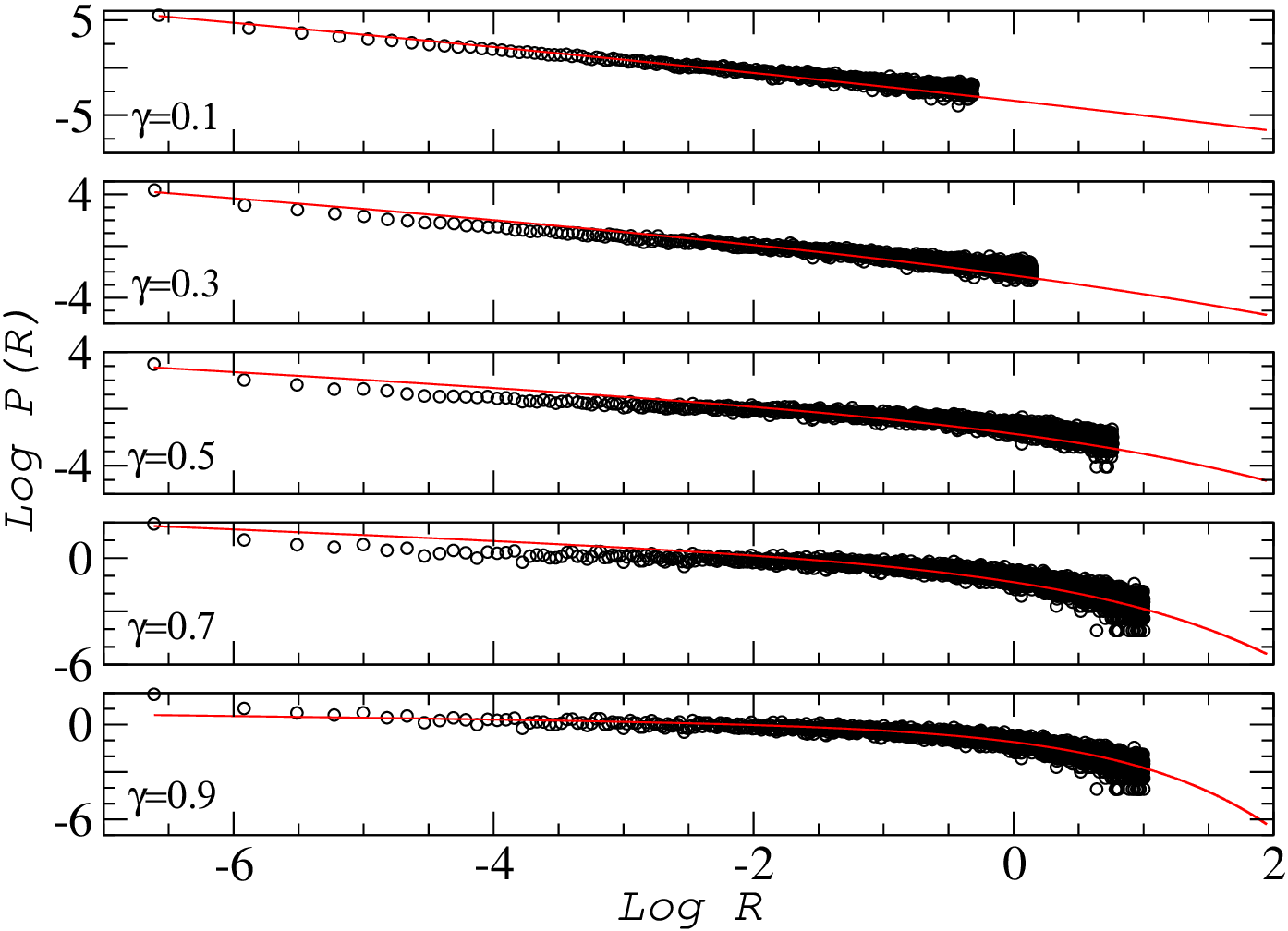}
\caption{(Color Online) The simulated return interval distribution (circles) and theoretical
distribution in Eq \ref{rtd4} (solid lines) for long range correlated time series. The threshold
is $q=3.0$ with average return interval $\langle r \rangle = 743.0$ for all
the cases shown above.}
\label{distrib1}
\end{figure}
\end{widetext}

Using Eq. \ref{rtd3} in practice can lead to strong divergence for $r \to 0$.
From a physical standpoint, this represents a problem that can be understood
based on the fact that
there cannot be zero return intervals, but they can be arbitrarily small.
By definition $r >0$, and if $r_{min}$ is the shortest return interval then its
corresponding scaled version would
be $r_{min}/\langle r \rangle $. If the original signal is sampled at equal time
intervals, $r_{min}$ can be scaled to unity and the shortest scaled return
interval would be $1/\langle r \rangle $.
The modification of Eq. \ref{rtd3} should be done by replacing the lower limit
in the integrals in Eqs \ref{normint0} and \ref{normint}
by $1/\langle r \rangle$
instead of 0. This also reflects the general idea that all power laws in
practice have a lower bound and the return
interval distribution like the Eq. \ref{rtd3} that displays
a power law type regime will necessarily have a lower cut off.

We will go back to Eqn. \ref{rtd2} and rewrite the return interval
distribution as
\begin{equation}
f(r) =  B ~r^{-(1-\gamma)} ~e^{-\frac{A}{\gamma} r^{\gamma}},
\label{rtd4}
\end{equation}
where $A$ and $B$ are constants that would now
depend on both $\gamma$ and the average return interval.
As usual, both these constants will be fixed by demanding that
probability and average return interval normalize to unity.
This leads to the following set of integrals;
\begin{equation}
\int_{s_0}^{\infty} ~f(r) ~dr = \frac{B}{A} e^{-p} = 1,
\label{renormint1}
\end{equation}
\begin{equation}
\int_{s_0}^{\infty} r ~f(r) ~dr = \frac{B s_0}{A} \left( e^{-p} + 
           \frac{\Gamma(1/\gamma,p)}{\gamma ~p^{1/\gamma}} \right) = 1
\label{renormint2}
\end{equation}
where $s_0=1/\langle r \rangle$, $p= A s_0^{\gamma}/\gamma$ and
$\Gamma(.,.)$ is the incomplete Gamma function \cite{igf}. The
algebraic equations to be solved for $A$ and $B$ are transcendental
in nature and closed form solution does not seem possible except
for some special values. By further manipulation of Eqns. \ref{renormint1} and
\ref{renormint2}, we obtain
\begin{equation}
\frac{1}{s_0} = 1 + \frac{e^p ~\Gamma(1/\gamma,p)}{\gamma ~p^{1/\gamma}}.
\label{trans1}
\end{equation}
If $p=p_0$ is the solution of Eq \ref{trans1} for a definite $\langle r \rangle$,
then the constants can be obtained as,
\begin{equation}
A = \frac{\gamma ~p_0}{s_0^\gamma}, \;\;\;\;\;\;\;\;\;\;\;\;\;\; B = A e^{p_0}.
\label{trans2}
\end{equation}
In the simulations shown in this paper, we have numerically solved for
constants $A$ and $B$ in Eq. \ref{rtd4} for various values of $\langle r \rangle$
using Eqns \ref{trans1} and \ref{trans2}.

\section{Numerical Results}
In this section, we display the numerical results for the return
interval distribution of long range correlated time series drawn
from a Gaussian distribution with zero mean and unit variance.
The long range correlated data was generated using the Fourier
filtering technique \cite{fft}. We generate $2^{25} \sim 3 \times 10^7$ data
points for each
values of $\gamma$ and then compute their return interval distribution.
The numerical results are displayed in Fig \ref{distrib1} as log-log plot
for $q=3$ along with the
theoretical distributions given in Eqns. \ref{rtd3} and \ref{rtd4}.
The agreement with the theoretical distribution is good and as expected
gets better as $\gamma \to 1$. Similar good agreement is also obtained
for the values of $\gamma$ not shown here. The simulated results in Fig \ref{distrib1}
does not cover a larger range in $\log R$ because of the large value of 
threshold $q$ chosen corresponding to an average return interval of
$\langle r \rangle = 743.0$. To over come this problem, we will need extremely
large sequences of random time series.
As we have argued in the previous section, the theoretical distribution
can be expected
to agree with the data when threshold $q$ or equivalently the
average return interval is large. Thus, as we reduce $q$ below 2.5,
there are deviations from the theoretical distribution which are systematically
studied in the next section.

\begin{figure}
\includegraphics[width=7.5cm]{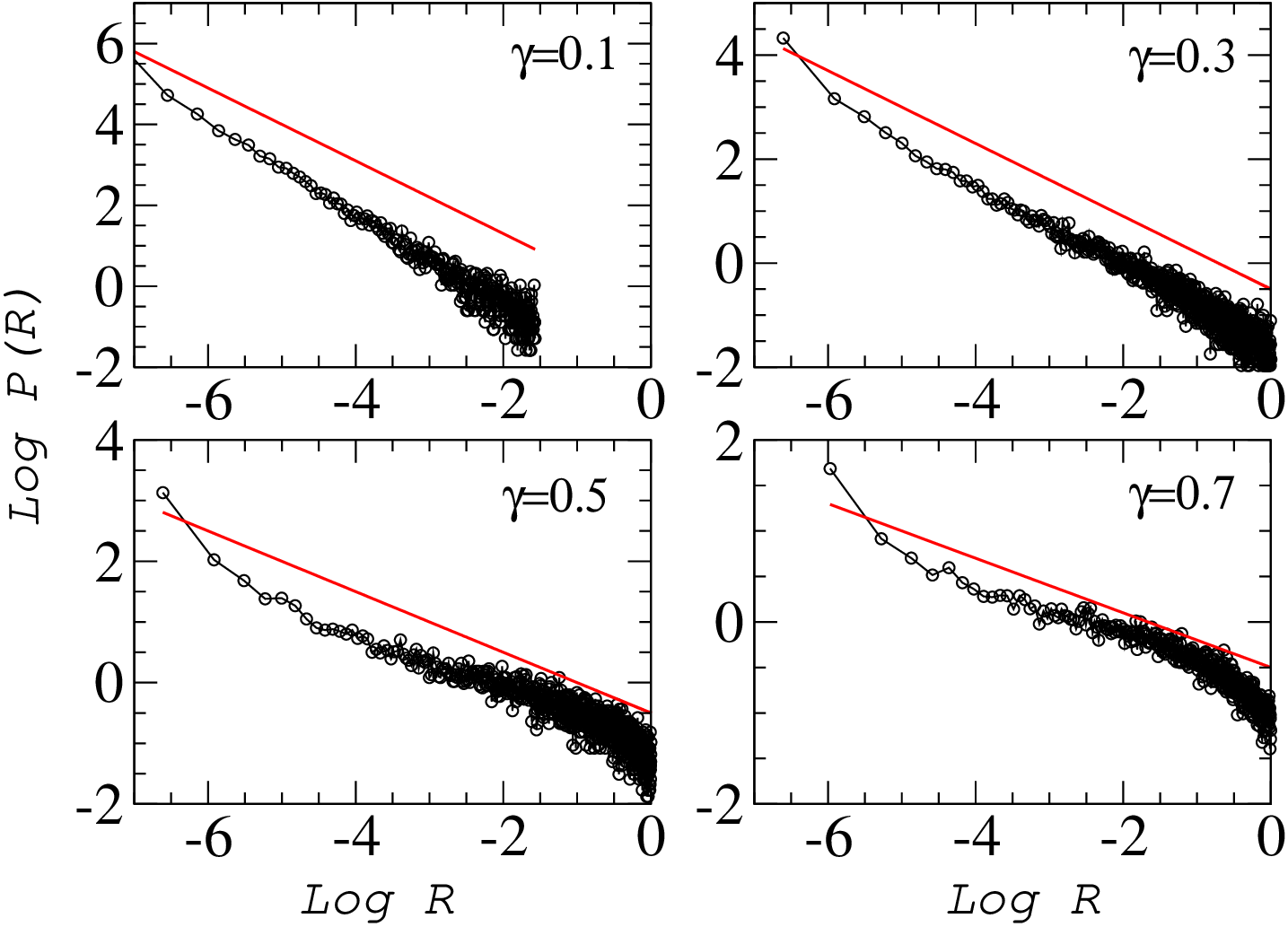}
\caption{(Color Online) The return interval distribution focused on the power law regime.
The numerical distribution (circles) is  nearly a straight line
with the slope $(-1+\gamma)$. A straight line with slope $(-1+\gamma)$
is shown as solid (red) line for comparison. For all the cases, $q > 3.0$
corresponding to $\langle r \rangle > 740.0$.}
\label{plaw}
\end{figure}

In Fig \ref{plaw}, we show the power law regime indicated by Eq. \ref{power}.
In this figure, we focus on the region $R \ll 1$ where we expect the power
law to appear. For each value of $\gamma$ in Fig \ref{plaw}, we have
drawn a straight line (shown in red) with the slope $(-1+\gamma)$. Quite
clearly, the numerical data show a remarkably good agreement with the
theoretical slope. As $\gamma\to 0$, the power law regime holds good in a larger
range of $R$ ; for instance, see the case of $\gamma=0.1 ~\mbox{and} ~0.3$. On the other
hand, as seen in the case of $\gamma=0.7$, the power law region becomes shorter
and stretched exponential regime begins to dominate
as $\gamma \to 1$. This is an indication that the return interval distribution
makes a transition from predominantly (stretched) exponential behavior to
predominantly power law type curve as  $\gamma \to 0$.
It must be pointed out that the agreement with theoretically expected slope
$(-1+\gamma)$ is reached only for $q \gg 1$. This is to be expected since the
derived distributions in Eqns \ref{rtd3} and \ref{rtd4} do not take
into account the correlations among the return intervals.
In the next section, we study how the slope in the power law regime
changes with threshold $q$ in the numerically simulated long range
correlated data.

\section{Return interval distribution and threshold for extreme events}
\begin{figure}[!ht]
\includegraphics*[width=8.2cm]{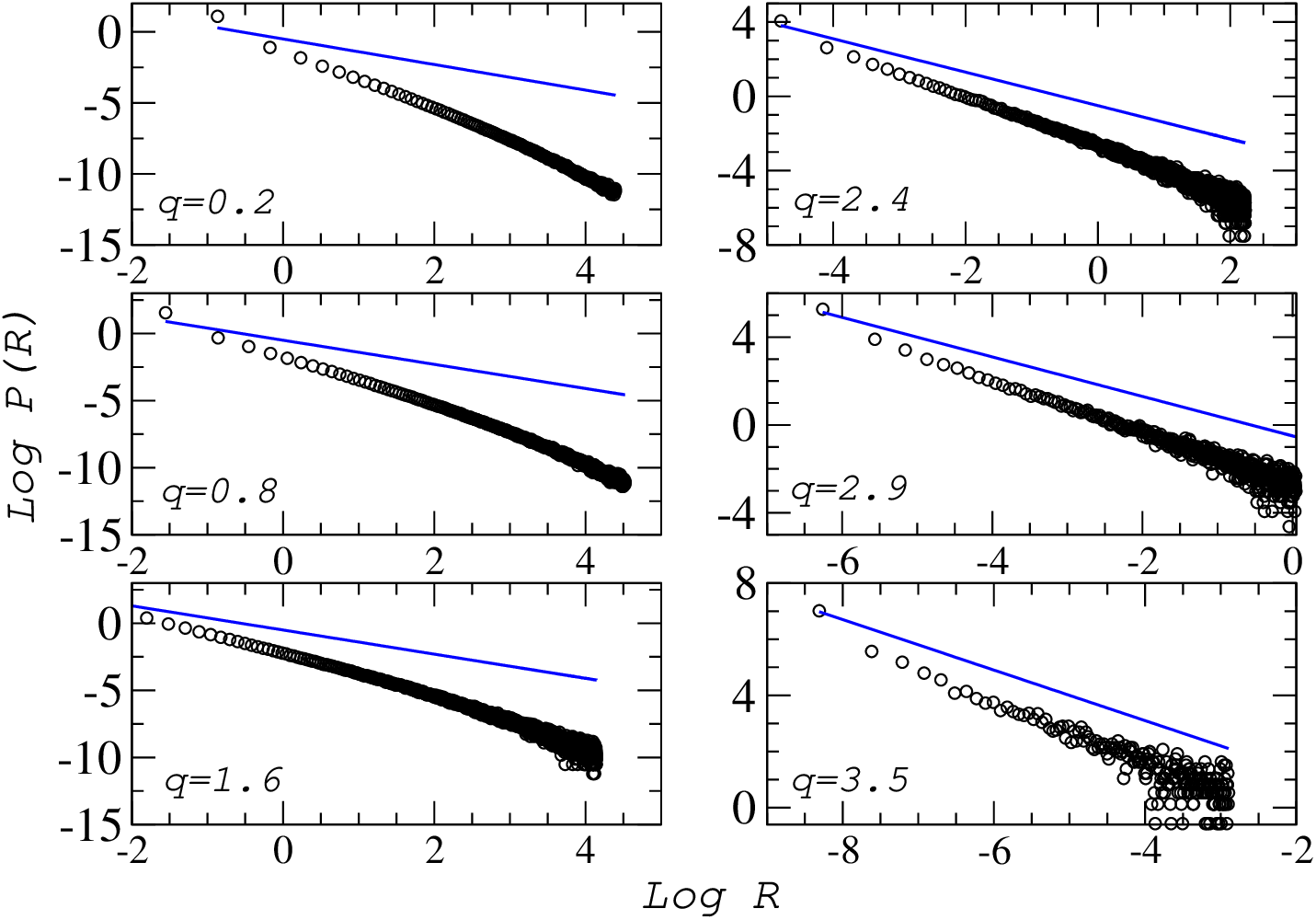}
\caption{(Color Online) The return interval distribution for the simulated data (circles) for
$\gamma=0.1$ plotted for various values of threshold $q$.
A straight line with slope $(-1+\gamma)$
is shown as solid (blue) line for comparison. Note that as $q$ increases,
the initial part of the distribution moves closer to a slope of $(-1+\gamma)$.}
\label{rtdvsq}
\end{figure}

In this section, we will empirically examine the relation between the
return interval distribution, especially in the power law regime, and
the threshold $q$ that define the extreme events.
Intuitively, we can expect that if the threshold is higher, extreme
events will be fewer and hence the return intervals will be longer.
Thus, larger $q$ leads to larger average return intervals.
Here we address the question of how the return interval distributions
in Eqns \ref{rtd3} and \ref{rtd4} are modified by changes in
threshold value $q$. One clear indication is that, approximately for $q<2$, the
simulated return interval distributions deviate systematically from 
Eqns \ref{rtd3} and \ref{rtd4}, in particular for $R<1$.
To study this, we plot the return interval distribution for the
simulated data in a log-log plot
as shown in Fig \ref{distrib1} and measure the slope in a linear region
for $R<1$ for various values of $q$. The result is displayed in Fig \ref{rtdvsq}
for $\gamma=0.1$. It is seen that as $q$ increases, the initial part of
the distribution, i.e, $R<1$ or $\log R < 0$, is closer to being a straight
line with slope $(-1+\gamma)$. A similar behavior is seen for all the values
of $\gamma$ of our interest.

\begin{figure}
\includegraphics*[width=8.2cm]{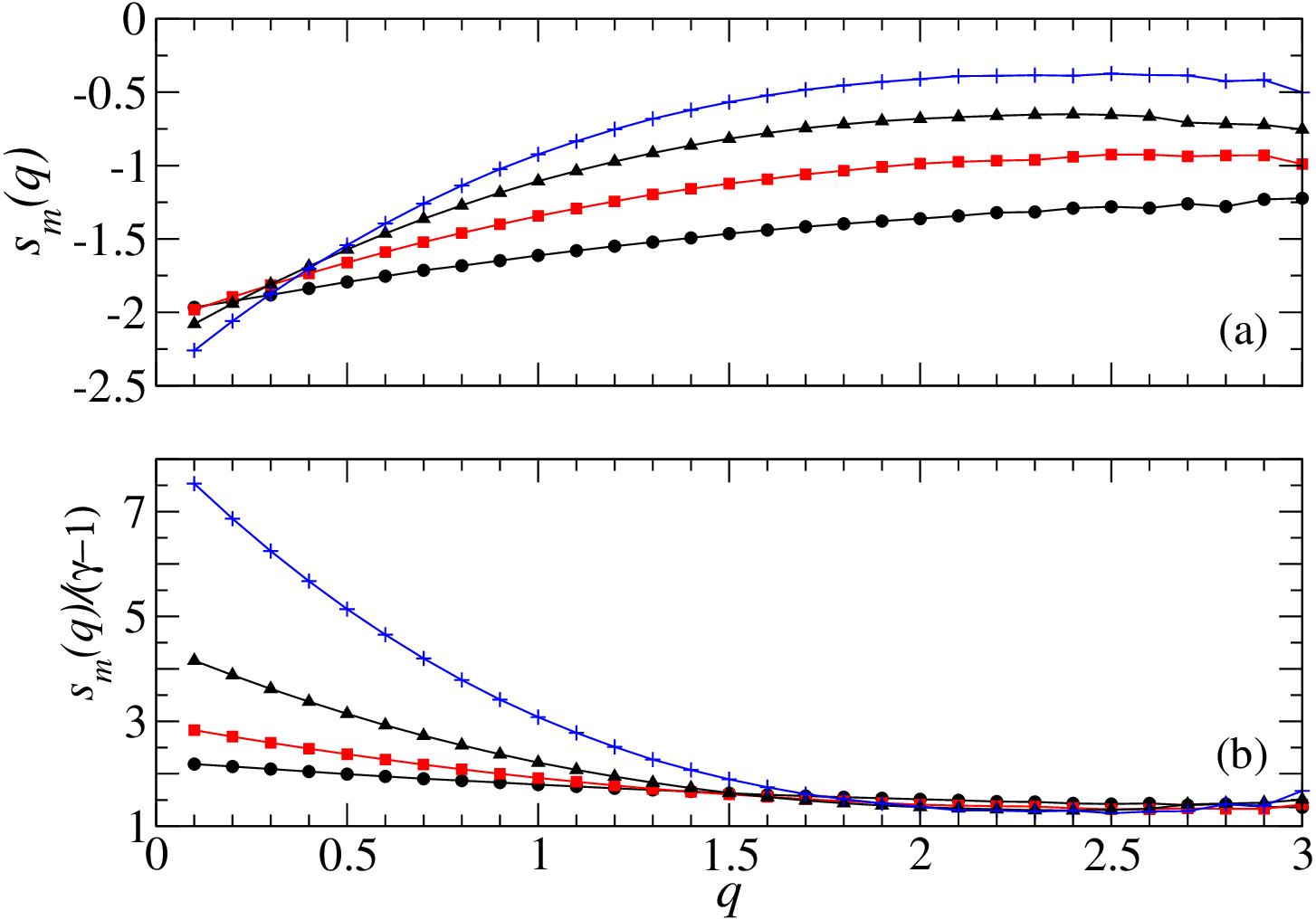}
\caption{(Color Online) (a) The measured slope $s_m$ in the power law regime  as a function
of $q$ for $\gamma=0.1$ (circles), 0.3 (squares), 0.5 (triangles) and
0.7 (plus). (b) The function $\theta(q,\gamma) = s_m/(\gamma-1)$ as a function of $q$
for same values of $\gamma$ as in (a).}
\label{slopeq}
\end{figure}

In order to see this variation of the slope of the initial part of
the distribution with $q$, we plot in Fig \ref{slopeq}(a) the measured slope $s_m(q)$
against the threshold $q$ for various values of $\gamma$. The slope is
measured in the linear region in log-log plot for $R \ll 1$. For a given
value of $\gamma = \gamma_c$, the slope increases monotonically
to reach a saturation value of $(-1+\gamma_c)$ as $q\to\infty$.
Once again we point out that this is in agreement with our expectation
that the weakly correlated regime would agree with the distribution
obtained in Eq. \ref{rtd3} and \ref{rtd4}. For the Gaussian
distributed data that we use, at $q=3$, the average
return interval is $\langle r \rangle \approx 744.0$.
Beyond $q=3$ with $2^{25}$ data points the
number of returns intervals are not sufficient for reliable statistics.
All this would imply that in order to take into account the effect of
$q$, the power law proposed in Eq. \ref{power} could be modified
as
\begin{equation}
P(R) \propto R^{-(1-\gamma) \theta(q,\gamma)},
\end{equation}
with the restriction, suggested by the numerical results in Fig \ref{slopeq}(a),
that $\theta(q,\gamma) \to 1$ as $q\to\infty$.
Clearly, the measured slope is simply given by $s_m = -(1-\gamma) \theta(q,\gamma)$.
Thus, we can directly visualize the function $\theta(q,\gamma)$ if we plot
$s_m(q)/(\gamma-1)$ as a function of $q$. This is shown in Fig \ref{slopeq}(b).
As we anticipated, the function $\theta(q,\gamma)$ tends towards unity as $q \to \infty$.
The autocorrelation exponent $\gamma$ controls the rate at which the
limiting value of unity is reached.
We believe that the behavior displayed in Fig \ref{slopeq}(a,b) is related to a more
fundamental question of how
the auto-correlation exponent of a long range correlated time series changes
if it is subjected
to thresholding such as the one we have applied using Eq. \ref{threshold}.
Obviously, every time we choose a subset of events from a larger set, such as the
extreme events, implicitly such thresholding is applied.
Since the power law regime varies with $q$ and if the distribution has to
remain normalized, then the stretched exponential part would also be modified.
However, this might be difficult to visualize numerically.
The central premise of this section is to show that Eqns \ref{rtd3} and \ref{rtd4}
represent return interval distributions in the limit when the threshold
or average return interval is large. We have shown through simulations the
dependence of return interval distributions on threshold $q$. This
explains why we have chosen $q=3$ to illustrate our result in Fig \ref{distrib1}.
Thus, in principle, the exact return interval distribution should depend
on $\langle r \rangle$, especially for short return intervals, i.e, $R < 1$.

\section{Long range probability process}
 Apart from corrections arising due to dependence on $q$, the return interval
distribution derived in this paper suffers due to approximation arising from
assumptions of independence of return intervals. This assumption makes the
analysis tractable but does not reflect the reality since we know that
the intervals are indeed correlated. In this section, we argue
that the deviations from the numerical simulations evident in Fig \ref{distrib1}
can be attributed to the presence of correlations in the return interval data.
We do this by simulating the probability process in Eq. \ref{probmodel} that forms the
basis for the analytical result in Eqn \ref{rtd3} and \ref{rtd4}.
If the simulated data agrees with the analytical result, then we could
attribute the deviations seen in Fig \ref{distrib1} to the correlations
present in the return intervals.

\begin{figure}
\includegraphics*[width=8.2cm]{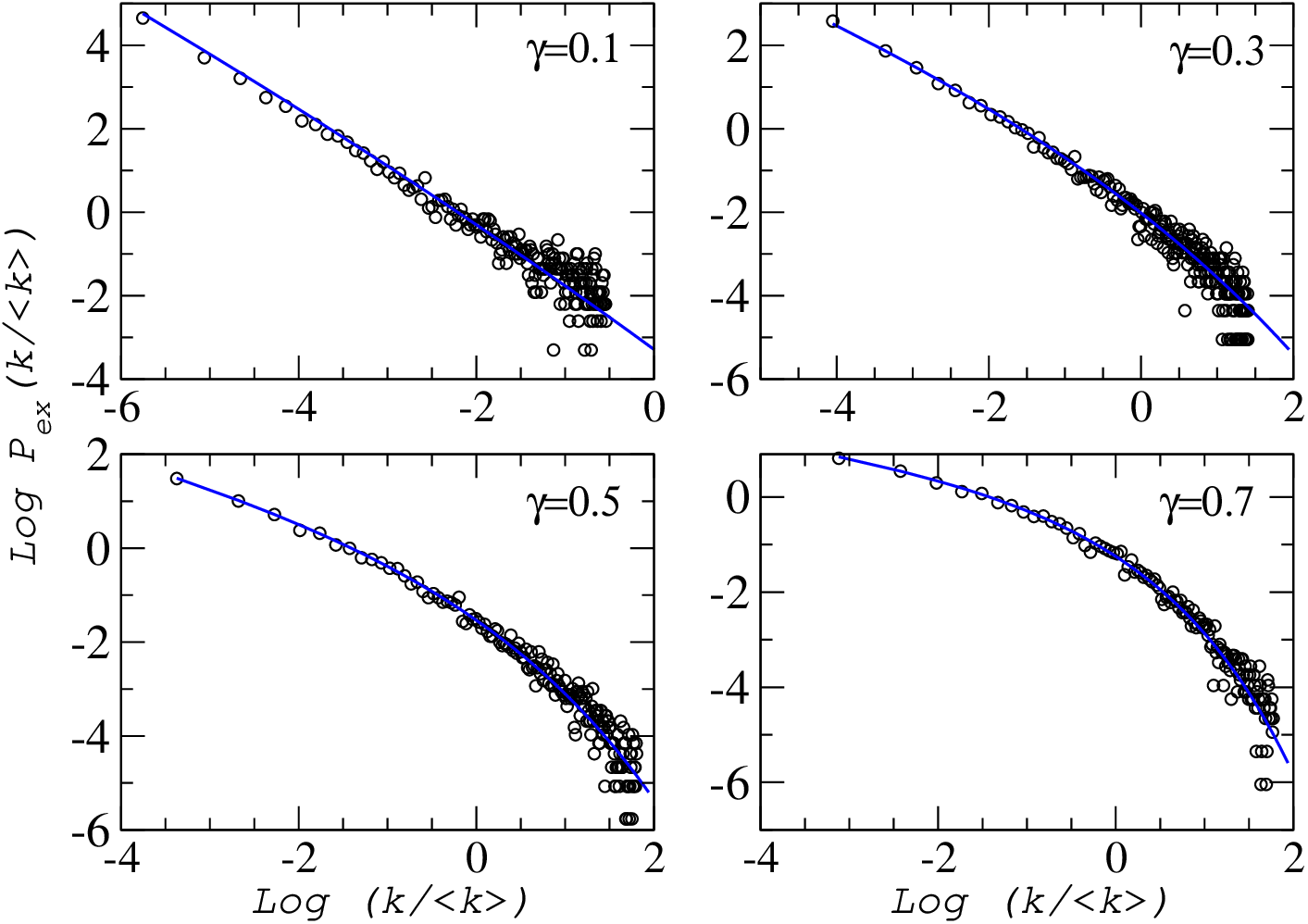}
\caption{(Color Online) The simulated return interval distribution (circles) from the probability
process in Eq. \ref{probmodel} compared with the theoretical
distribution (solid line) given in Eq \ref{rtd4}.}
\label{pmodel}
\end{figure}

In order to numerically simulate the probability process in Eq. \ref{probmodel},
we first determine the constant $a$ by normalizing it in the region $k_{min}=1$ and
$k_{max}$. The normalized probability distribution corresponding
to Eq. \ref{probmodel} is
\begin{equation}
P(k) = \frac{\gamma}{ \left( k_{max}^{\gamma} - 1 \right) } ~k^{-\gamma},
\end{equation}
where $k=1,2,3....$. We generate a random number $\xi_k$ from a uniform distribution
at every $k$ and compare it with the value of $P(k)$. A random number
is accepted as an extreme event
if $\xi_k < P(k)$ at any given value of $k$. If $\xi_k \ge P(k)$, then
it is not an extreme event. By this procedure, we generate a series of
extreme events following Eq. \ref{probmodel}. We then compute the return
intervals and its distribution after scaling it by the average return interval.
In Fig. \ref{pmodel}, we show the return interval distribution obtained
by simulating our probability process along with the distribution given
by Eq \ref{rtd4}. The agreement with the theoretical distribution is
excellent, including for the values of $\gamma$ not shown here.
Hence, if the long range correlated data had independent
return intervals, then we would have obtained nearly perfect agreement
with Eq \ref{rtd3} and \ref{rtd4}. This implies that the remaining disagreement
between the theoretical and numerical results seen in Fig \ref{distrib1} can
be attributed to the
presence of correlations among the return intervals. On the other hand,
if the probability process in Eq \ref{probmodel} was an incorrect assumption,
it may not have been possible to obtain the results displayed in Fig \ref{distrib1}.

\section{Discussions and conclusions}
  We have studied the distribution of return intervals for the extreme
events in long range correlated time series. 
An approximate analytical expression for this distribution has been
obtained starting from the empirically established fact that
returns intervals are long range correlated. This distribution
is a product of a power law and a stretched exponential and explains
the observed power law for short return intervals. 
For large return
intervals, the distribution is dominated by a stretched exponential
decay. The works reported earlier have empirically proposed
stretched exponential form for the return interval distribution
which is now shown to be valid in the domain of large return
intervals. Further, we have also carefully studied the role
played by the threshold $q$ or equivalently the average return
interval in the return time statistics. We show that it modifies
the return interval distribution, especially in the power law
regime of short return intervals. We believe that the results
obtained in this paper explains most of the empirically observed
features in the return time distributions of long range correlated time series.
In the simulations reported in this work, we have used Gaussian
distributed random numbers. As studied in Ref. \cite{long3}, it
is natural to ask if the exponential or power law distributed data
would modify the results of this paper. We expect that the functional
form of the distribution in Eq \ref{rtd4} would not be modified though the
normalization constants $A$ and $B$ might change due to their
dependence on the threshold $q$. The question of verifying the
results of this paper with a measured time series is underway
and would be reported elsewhere.

 As pointed out before, the inter-event time distribution has
applications across many disciplines. Hence, it appears in different
settings in different areas. In the statistical literature,
a related problem of zero crossings, i.e, the probability that $X(t)>0$
for $0 \ge t \ge T$ has been considered. Under certain conditions, 
for a stationary Gaussian process, the upper bound for zero crossing
probability is shown to be a
stretched exponential \cite{newell}. This result does not strictly
apply to the case of recurrence interval statistics because the
zero crossing probability does not make statements about occurrence or
non-occurance of another zero crossing after the interval $T$.
A return interval, by definition, requires two crossings separated by
an interval with no crossings.
Finally, we would like to remark that the analytical distribution obtained
in this paper appears to be related to the universal scaling form
proposed recently \cite{corral} in the context of earth quakes but
appears to be more generally valid. Thus it is likely that the
exact return interval distribution might incorporate corrections to
the one obtained in this paper. Indeed, if the exact distribution
is known,
it will also become possible to determine the precise time scales
over which power law and exponential decay operate. This, in turn,
should help address questions of hazard estimation for extreme
events more carefully and, needless to say, this has enormous
interest in the insurance industry \cite{reiss} and as a tool for
decision support system \cite{dss}.

\end{document}